\documentclass[pra,showpacs,superscriptaddress]{revtex4}

\usepackage{bm}

\input{epsf}

\begin{document}
\draft
\title{Entanglement, Mixedness, and Spin-Flip Symmetry in Multiple-Qubit Systems\\}

\author{Gregg~Jaeger}

\affiliation {Quantum Imaging Laboratory, Department of Electrical
and Computer Engineering, Boston University, Boston, MA 02215}

\author{Alexander~V.~Sergienko}

\affiliation {Quantum Imaging Laboratory, Department of Electrical
and Computer Engineering, Boston University, Boston, MA 02215}
\affiliation{Department of Physics, Boston University, Boston, MA
02215}

\author{Bahaa~E.~A.~Saleh}

\affiliation {Quantum Imaging Laboratory, Department of Electrical
and Computer Engineering, Boston University, Boston, MA 02215}

\author{Malvin~C.~Teich}

\affiliation {Quantum Imaging Laboratory, Department of Electrical
and Computer Engineering, Boston University, Boston, MA 02215}
\affiliation{Department of Physics, Boston University, Boston, MA
02215}

\date{\today}

\renewcommand\baselinestretch{2}\small\normalsize
\begin{abstract}

A relationship between a recently introduced multipartite
entanglement measure, state mixedness, and spin-flip symmetry is
established for any finite number of qubits. It is also shown
that, for those classes of states invariant under the spin-flip
transformation, there is a complementarity relation between
multipartite entanglement and mixedness. A number of example
classes of multiple-qubit systems are studied in light of this
relationship.

\end{abstract}
 \maketitle

\section{Introduction.}

State entanglement and mixedness are properties central to quantum
information theory. It is therefore important, wherever possible,
to relate them. The relationship between these quantities has been
previously investigated  for the simplest case, that of two-qubits
(see, for example, \cite{W1, Metal, Wetal, Getal}), but has been
much less well studied for multiple-qubit states because
multipartite entanglement measures (\cite{Betal, Thap}) have only
been recently given in explicit form (see, for example, \cite{EB,
Jetal, Miyake, WC}). Here, we find a relationship, for any finite
number of qubits, between a multipartite entanglement measure and
state mixedness, through the introduction of a multiple-qubit
measure of symmetry under the spin-flip transformation. Here,
several classes of two, three and four qubit states are examined
to illustrate this relationship. Since quantum decoherence affects
the mixedness and entanglement properties of multiple-qubit
states, our results provide a tool for investigating decoherence
phenomena in quantum information processing applications, which
utilize just such states.

The purity, $\mathcal{P}$, of a general quantum state can be given
in all cases by the trace of the square of the density matrix,
$\rho$, and the mixedness, $M$, by its complement:
\begin{eqnarray}
\mathcal{P}(\rho)&=&{\rm Tr}\,\rho^2\ ,\nonumber\\
M(\rho)&=&1-{\rm Tr}\,\rho^2=1-\mathcal{P}(\rho)\ .
\end{eqnarray}
Entanglement can be captured in several ways, which may or may not
be directly related, with varying degrees of coarseness (see, for
example, \cite{EB}), depending on the complexity of the system
described (see, for example, \cite{Wootters, Zs, ASST, GSV}).
Here, we will measure entanglement by a recently introduced
measure of multipartite entanglement, the SL(2,C)$^{\times
n}$-invariant quantity
\begin{equation}
S_{(n)}^2={\rm Tr}(\rho\tilde{\rho})\ ,
\end{equation}
where the tilde indicates the spin-flip operation \cite{Jetal}.
The multiple qubit spin-flip operation is defined as
\begin{eqnarray}
\rho\longrightarrow\tilde{\rho}\equiv\sigma_2^{\otimes n}
\rho^*\sigma_2^{\otimes n}\ ,
\end{eqnarray}
where $\rho^*$ is the complex conjugate of the $n$-qubit density
matrix, $\rho$, and $\sigma_2$ is the spin-flipping Pauli matrix
\cite{U}.  Its connection to the $n$-tangle, another multipartite
entanglement measure, will be discussed below. We now discuss this
measure in relation to the well-established, bipartite measures of
entanglement \cite{W1}.

In the simplest case of pure states of two qubits (A and B),
entanglement has most commonly been described by the entropy of
either of the one-qubit reduced density operators, which is
obtained by tracing out the variables of one or the other qubits
from the total system state described by the projector
$P[|\Psi_{AB}\rangle]\equiv|\Psi_{AB}\rangle\langle\Psi_{AB}|$.
For mixed two-qubit states, $\rho_{AB}$,  the entanglement of
formation, $E_f$, is given by the minimum average marginal entropy
of all possible decompositions of the state as a mixture of
subensembles. Alternatively, one can use a simpler measure of
entanglement, the concurrence, $C$, to describe two-qubit
entanglement \cite{Wootters}. For pure states, this quantity can
be written
\begin{equation}
C(|\Psi_{AB}\rangle)=
|\langle\Psi_{AB}|\sigma_2\otimes\sigma_2|\Psi_{AB}^*\rangle|
=|\langle\Psi_{AB}|\tilde{\Psi}_{AB}\rangle|\ ,
\end{equation}
where $|\tilde{\Psi}_{AB}\rangle$ is the spin-flipped state
vector. It has been shown that the concurrence of a mixed
two-qubit state, $C(\rho_{AB})$, can be expressed in terms of the
minimum average pure-state concurrence, $C(|\Psi_{AB}\rangle)$,
where the minimum is taken over all possible ensemble
decompositions of $\rho_{AB}$ and that, in general, $C(\rho)={\rm
max}\{0, \lambda_1-\lambda_2-\lambda_3-\lambda_4\}$, where the
$\lambda_i$ are the square roots of the eigenvalues of the product
matrix $\rho\tilde{\rho}$, the ``singular values,'' all of which
are non-negative real quantities \cite{Wootters}. It has also been
shown that the entanglement of formation of a mixed state $\rho$
of two qubits can be expressed in terms of the concurrence as
\begin{equation}
E_f(\rho)=h(C(\rho))\ ,
\end{equation}
where $h(x)=-x\ {\rm log}_2x-(1-x)\ {\rm log}_2(1-x)$
\cite{Wootters}.

Here, we measure multipartite entanglement involving $n$ qubits by
$S_{(n)}^2$\ (see \cite{Jetal}), which is invariant under the
group corresponding to stochastic local operations and classical
communications (SLOCC) \cite{LO}. For two-qubit pure states, this
measure (with $n=2$) coincides with the squared concurrence (or
tangle):
\begin{equation}
S_{(2)}^2(P[|\psi\rangle])=\tau_{(2)}(P[|\psi\rangle])=C^2(P[|\psi\rangle])\
,
\end{equation}
where $P[|\psi\rangle]\equiv|\psi\rangle\langle\psi|$ is the
projector corresponding to its state-vector argument,
$|\psi\rangle$.

In Section II, after providing another expression for $S_{(n)}^2$
for all values of $n$ for both pure and mixed states, we show that
for pure states this length coincides with the multipartite
generalization of the tangle defined for pure states of any finite
number of qubits, $n$, that is, the pure state $n$-tangle
\cite{WC, Miyake}. We discuss the geometrical properties of
$S_{(n)}^2$ and of the purity, $\mathcal{P}$, that allow us to
find the general relationship between multipartite entanglement
and mixedness in terms of a spin-flip symmetry measure. In
Sections III and IV, we examine a range of classes of two-,
three-, and four-qubit states to illustrate the results of the
Section II.

\section{Definitions and the General Case}

The multiple-qubit SLOCC invariant given in Eq. 2 is most
naturally defined in terms of $n$-qubit Stokes parameters
\cite{Fano, FM},
\begin{equation}\label{Stensor}
  S_{i_1...i_n}= \mathrm{Tr} (\rho\ \sigma_{i_1}\otimes
  \cdots\otimes\sigma_{i_n})\ , \ \ i_1, ...,i_n=0,1,2,3\ ,
\end{equation}
where $\sigma_\mu^2=1$, $\mu = 0, 1, 2, 3$, are the three Pauli
matrices together with the identity $\sigma_0=I_{2\times 2}$, and
${1\over 2}{\rm Tr}(\sigma_\mu\sigma_\nu) =\delta_{\mu\nu}$. These
directly observable parameters form an $n$-particle generalized
Stokes tensor $\{S_{i_1 ...i_n}\}$. Under SLOCC transformations
\cite{Betal}, density matrices undergo local transformations
described by the group SL(2,C), while the corresponding Stokes
parameters undergo local transformations described by the
isomorphic group O$_0(1,3)$, the proper Lorentz group
\cite{Sternberg}, that leave the Minkowskian length unchanged. The
Minkowskian squared-norm of the Stokes tensor $\{S_{i_1... i_n}\}$
provides this invariant length \cite{Jetal} (here renormalized by
the factor $2^{-n}$ for convenience):
\begin{eqnarray}\label{Invar}
  S_{(n)}^2&\equiv &{1\over 2^n}\bigg\{(S_{0 ...0})^2
-\sum_{k=1}^n\sum_{i_k=1}^3(S_{0 ...i_k ...0})^2 \
        \nonumber\\
&&\ \ \ \ \ \ \ +\sum_{k,l=1}^n\, \sum_{i_k, i_l=1}^3(S_{0 ...i_k
...i_l ...0})^2 - \cdots
        \nonumber\\
&&\ \ \ \ \ \ \ +\ (-1)^n \sum_{i_1,...,i_n=1}^3(S_{i_1 ...
i_n})^2\bigg\} .
\end{eqnarray}
As mentioned above, this quantity can be compactly expressed in
terms of density matrices, as
\begin{equation}
S^2_{(n)}=\mathrm{Tr} (\rho_{12... n}\, \tilde{\rho}_{12... n})\ ,
\end{equation}
\noindent where ${\rho}_{12... n}$ is the multiple-qubit density
matrix and $\tilde{\rho}_{1... n}=(\sigma_2^{\otimes n})\rho_{1...
n}^*(\sigma_2^{\otimes n})$ is the spin-flipped multiple-qubit
density matrix. Here, we consider these positive, Hermitian,
trace-one matrices as operators acting in the Hilbert space
$(\mathcal{C}^2)^{\otimes n}$ of $n$ qubits. For the remainder of
this article, subscripts on these matrices either will be
suppressed or replaced by more descriptive labels, the number $n$
of qubits being otherwise specified.

Like this measure of entanglement, the state purity,
$\mathcal{P}$, is also naturally captured in terms of the
generalized Stokes parameters. The subgroup of deterministic local
operations and classical communications (LOCC) \cite{LOCC} on
qubits, namely the unitary group [SU(2)] of transformations on
density matrices, corresponds to the subgroup of ordinary
rotations [SO(3)] of Stokes parameters, that preserve the
Euclidean length derived from these parameters. The purity for a
general $n$-qubit state is the Euclidean length in the space of
multiple-qubit Stokes parameters:
\begin{equation}
\mathcal{P}(\rho)=\mathrm{Tr}\, {\rho^2}= {1\over 2^n}\sum_{i_1,
..., i_n=0}^3S^2_{i_1 ...i_n}\ ,
\end{equation}
(see \cite{Jetal}).

The multiple-qubit state purity  (Eq. 10) and the entanglement
(Eqs. 8, 9) can be related by the (renormalized) Hilbert-Schmidt
distance in the space of density matrices (arising from the
Frobenius norm \cite{RS}):
\begin{equation}
D_{\rm HS}(\rho-\rho')\equiv \sqrt{\frac{1}{2}{\rm
Tr}[(\rho-\rho')^2]}\ .
\end{equation}
In particular, the multi-partite entanglement and mixedness are
related by the square of the Hilbert-Schmidt distance between the
state $\rho$ and its corresponding spin-flipped counterpart
$\tilde{\rho}$:
\begin{eqnarray}
D_{\rm HS}^2(\rho-{\tilde\rho})&=&\frac{1}{2}{\rm Tr}[(\rho-{\tilde\rho})^2]\nonumber\\
&=&\frac{1}{2}\Big[{\rm Tr}\,\rho^2+{\rm Tr}\,{\tilde \rho}^2-2
{\rm Tr}(\rho{\tilde\rho})\Big]\nonumber\\
&=&{\rm Tr}\,\rho^2-{\rm Tr}(\rho{\tilde\rho})\nonumber\\
&=&\mathcal{P}(\rho)-S_{n}^2(\rho)\ .
\end{eqnarray}
Thus, we have the following relation between the chosen measure of
multi-partite state entanglement and the state purity:
\begin{equation}\label{comp0}
S_{n}^2(\rho)+D_{\rm HS}^2(\rho-\tilde{\rho})=\mathcal{P}(\rho)\ ,
\end{equation}
where $D_{\rm HS}^2(\rho-\tilde{\rho})$ can be understood as a
measure of distinguishability between the $n$-qubit state $\rho$
and the corresponding spin-flipped state $\tilde{\rho}$ (as
defined in Eq. 3).

The relation of Eq. (13) can be recast as the following simple,
entirely general, relation between multipartite entanglement,
$S_{(n)}^2(\rho)$, and mixedness, $M(\rho)=1-\mathcal{P}(\rho)$:
\begin{equation}\label{comp}
S_{(n)}^2(\rho)+M(\rho)=I(\rho, {\tilde\rho})\ ,
\end{equation}
where we have introduced $I(\rho,{\tilde\rho})\equiv 1-D_{\rm
HS}^2(\rho-{\tilde\rho})$ which measures the indistinguishability
of the density matrix, $\rho$, from the corresponding spin-flipped
state, $\tilde{\rho}$; $I(\rho,\tilde{\rho})$ is clearly also a
measure of the spin-flip symmetry of the state.

For pure states, $M(\rho)=M(|\psi\rangle\langle\psi|)=0$, and we
have
\begin{equation}
S_{(n)}^2(|\psi\rangle\langle\psi|)=I(|\psi\rangle\langle\psi|,\,
|\tilde{\psi}\rangle\langle\tilde{\psi}|)\ .
\end{equation}
In this case, the Minkowskian length, $S_{(n)}^2(\rho)$, is also
seen to be equal to the $n$-tangle, $\tau_{(n)}$, as mentioned
above:
\begin{eqnarray}
S_{(n)}^2(|\psi\rangle\langle\psi|)&=&{\rm Tr}[(|\psi\rangle\langle\psi|)(|\tilde{\psi}\rangle\langle\tilde{\psi}|)]\nonumber\\
&=&\langle\psi|(|\tilde{\psi}\rangle\langle\tilde{\psi}|)|\psi\rangle\nonumber\\
&=&|\langle\psi|\tilde{\psi}\rangle|^2\nonumber\\
&=&\tau_{(n)}\ ,
\end{eqnarray}
where $|{\tilde\psi}\rangle\equiv\sigma_2^{\otimes
n}|\psi^*\rangle$ and
$\tau_{(n)}=|\langle\psi|\tilde{\psi}\rangle|^2$ is the pure state
$n$-tangle. Thus, we also see that for pure states
\begin{equation}
\tau_{(n)}=I(\rho,\tilde{\rho})\ ,
\end{equation}
that is, the $n$-tangle coincides with the degree of spin-flip
symmetry.

From Eq. (14), one obtains an exact complementarity relation for
those classes of n-qubit states, whether pure or mixed, for which
$\rho={\tilde\rho}$ (and thus $I(\rho,{\tilde\rho})=1$):
\begin{equation}\label{strict}
S_{(n)}^2(\rho)+M(\rho)=1\ ,
\end{equation}
since then $D_{\rm HS}(\rho -{\tilde\rho})=0$. For the special
case of pure states $M(\rho)=0$; this result can be understood as
expressing the fact, familiar from the set of Bell states, that
pure states of more than one qubit invariant under $n$-qubit
spin-flipping have full $n$-qubit entanglement. The Bell state
$|\Psi^+\rangle$ is the most obvious example from this class. By
contrast, for those states $\rho$ that are fully {\it
distinguishable} from their spin-flipped counterparts,
$\tilde{\rho}$, {\it i.e.} for which $I(\rho,\tilde{\rho})=0$, one
finds instead
\begin{equation}\label{distinguishable}
S_{(n)}^2(\rho)+M(\rho)=0\ ,
\end{equation}
implying that {\it both} the entanglement {\it and} the mixedness
are zero, $S_{(n)}^2=0$ and $M(\rho)=0$, since both quantities are
non-negative and sum to zero. An arbitrary $n$-fold tensor product
of states $|0\rangle$ and $|1\rangle$ is an $n$-qubit example from
this class. A third noteworthy case is that when the mixedness and
indistinguishability are non-zero and equal:
$M(\rho)=I(\rho,\tilde{\rho})$. In that case the entanglement
$S_{(n)}^2$ is obviously zero. The fully mixed $n$-qubit state
(described by identity matrix normalized to trace unity) is an
example from this last class.

\section{Two-qubit systems}

Relations between entanglement and mixedness have previously been
found for limited classes of two-qubit states using the tangle,
$\tau_{(2)}$, as an entanglement measure \cite{Wetal, Metal,
Getal}. In particular, for two important classes of states, the
Werner states and the maximally entangled mixed states, it was
found analytically that as mixedness increases, entanglement
decreases \cite{Wetal}. By exploring more of the Hilbert space of
two-qubit systems by numerical methods, it was also found that a
range of other states exceed the ratio of entanglement to
mixedness present in the class of Werner states including, in
particular, the maximally entangled mixed states. We now use our
results above to provide further insight into the relationship
between entanglement and mixedness in two-qubit systems, before
going on to examine larger multi-qubit systems.

Mixtures of two Bell states,
\begin{equation}
\rho_2=wP[|\Phi^+\rangle]+(1-w)P[|\Phi^-\rangle]\ ,\nonumber\\
\end{equation}
where $w\in[0,1]$, and mixtures of three Bell states,
\begin{equation}
\rho_3=w_1P[|\Phi^+\rangle]+w_2P[|\Phi^-\rangle]+w_3P[|\Psi^+\rangle]\
,
\end{equation}
where $w_i\in[0,1]$ and $w_1+w_2+w_3=1$, were both considered in
\cite{Getal}. Since both cases fall within the same larger class,
the Bell-decomposable states
\begin{eqnarray}
\rho_{BD}&=&w_1P[|\Phi^+\rangle]+w_2P[|\Phi^-\rangle]+\nonumber\\&&\
\ \ \ w_3P[|\Psi^+\rangle]+ w_4P[|\Psi^-\rangle]\ ,
\end{eqnarray}
where $w_i\in[0,1]$ and $w_1+w_2+w_3+w_4=1$, we consider them via
this general case. For these states, the general relation, Eq.
(14), tells us that, within this class, as the entanglement
increases the mixedness must decrease, as expressed by Eq. (18),
since all of these states are symmetric under the spin-flip
operation: $\rho={\tilde\rho}$. That is,
$M(\rho_{BD})=1-S_{(n)}^2(\rho_{BD})$. These states can be studied
in particular cases by performing Bell-state analysis, say on
qubit pairs within pure states of three particles, and can
characterize, for example, the ensemble of states transmitted
during a session of quantum communication using quantum dense
coding. The Werner states \cite{Werner} such as
\begin{equation}
\rho_{\rm Werner}=wP[|\Phi^+\rangle ]+{{1-w}\over 4}I_2\otimes
I_2\ ,
\end{equation}
which describe weighted mixtures of Bell states and fully mixed
states, where $w\in[0,1]$, also have the spin-flip symmetry
property, that is $I(\rho_{\rm Werner},\tilde{\rho}_{\rm
Werner})=1$. Because of their symmetry, all the above examples
obey the exact complementarity relation, Eq. (18): entanglement
and mixedness are seen to be strictly complementary. However, not
all two-qubit states possess full symmetry under spin flips.
Therefore, only the full three-way relation (Eq. 14) will hold in
general.

Let us now proceed further by considering two example classes
where the state is {\it not} invariant under the spin-flip
operation, one class of pure states and one class of mixed states.
For the two-qubit generalized Schr\"odinger cat states of the form
\begin{equation}
|\Phi (\alpha)\rangle_{AB}=\alpha
|00\rangle+\sqrt{1-\alpha^2}\,|11\rangle\ ,
\end{equation}
where $\alpha\in [0,1]$, considered in \cite{Getal}, clearly
$\tilde{\rho}\neq\rho$ in general. For them, one has for the state
distinguishability $D_{\rm HS}(\rho-{\tilde\rho})=|2\alpha^2-1|$,
and for the indistinguishability, $I(\rho,{\tilde\rho})=1-D_{\rm
HS}^2=4\alpha^2(1-\alpha^2)$. Since these states are pure, the
mixedness $M=0$ and the relation given by Eq. (14) (with $n=2$)
reduces to an expression for state entanglement (Eq. 15), that is
\begin{equation}
S_{(2)}^2(\rho)=\tau_{(2)}=4\alpha^2(1-\alpha^2)\ ,
\end{equation}
in accordance with Eq. (17). This shows the entanglement (in this
case coinciding with the tangle, $\tau_{(2)}$) to be parameterized
by $\alpha^2$, reaching its maximum at the maximally entangled
pure state of this class, $|\Psi^+\rangle$, as it should. We will
see below that similar expressions obtain for larger integral-$n$
Schr\"odinger cat states.

Now consider the class of two-qubit ``maximally entangled mixed
states,'' as defined in Ref. (3) as those states possessing the
greatest possible amount of entanglement among those states having
a given degree of purity. These can be written
\begin{eqnarray}
\rho_{\rm
mems}(\gamma)&=&\frac{1}{2}\big[2g(\gamma)+\gamma\big]P[|\Phi^+\rangle]
+\nonumber\\&&\ \ \ \
 \frac{1}{2}\big[2g(\gamma)-
 \gamma\big]P[|\Phi^-\rangle] +\nonumber\\&&\ \ \ \ \ \ \ \
\big[1-2g(\gamma)\big]P[|01\rangle]\ ,
\end{eqnarray}
where $g(\gamma)=\textstyle{\frac{1}{3}}$ for $\frac{2}{3}>\gamma
> 0$ and $g(\gamma)=\gamma/ 2$ for $\gamma\geq \frac{2}{3}$ \cite{Metal}.
For these states, we find that $I(\rho,\tilde{\rho})=1-D_{\rm
HS}^2(\rho-\tilde{\rho})=4g(\gamma)[1-g(\gamma)]$, so that
\begin{equation}
S_{(2)}^2(\rho_{\rm mems})+M(\rho_{\rm
mems})=4g(\gamma)\big[1-g(\gamma)\big]\ .
\end{equation}
The formal similarity of the expressions for degree of symmetry
$I(\rho,\tilde{\rho})$ in the above two cases, the generalized
Schr\"odinger cat and maximally entangled mixed states, is
noteworthy, with the degree of spin-flip symmetry of the state
being of the same form but with different arguments, $\alpha^2$
and $g(\gamma)$, respectively. This suggests that the maximally
entangled mixed states of Eq. (26) could substitute for
Schr\"odinger cat states in situations requiring such entanglement
and symmetry but where decoherence is unavoidable.

The above example classes have previously been used to explore the
relationship between entanglement and mixedness of states for a
given ability to violate the Bell inequality \cite{Metal,Getal}.
The behavior of these properties for Bell-decomposable states was
previously shown \cite{Getal} to differ in that context from that
of the Werner and maximally entanglement mixed states
\cite{Metal}. Therefore, no general conclusion could be drawn
about the relationship of entanglement and mixedness for the
general two-qubit state for a given amount of Bell inequality
violation. That is presumably because Bell inequality violation
indicates a deviation from classical behavior (see, for example,
\cite{RW}), rather than an inherent property of the state. The
results here show instead that the degree of state symmetry
governs the relationship of entanglement and mixedness.

\section{Three-qubit and four-qubit systems}

Let us first use the relation of Eq. (14) with $n=3$ to study
three-qubit states. For pure states, this relation takes the
special form given in Eq. (15) and provides the entanglement
measure $S_{(3)}^2$ directly in terms of the spin-flip symmetry
measure $I(\rho,\tilde{\rho})$. For the generalized W state with
complex amplitudes
\begin{equation}
|{\rm W}^g\rangle=\alpha|100\rangle+\beta |010\rangle
+\gamma|001\rangle\ ,
\end{equation}
with $|\alpha|^2+|\beta|^2+|\gamma|^2=1,$ using Eq. (15), we have
\begin{equation}
S_{(3)}^2={\rm Tr}(\rho_{{\rm W}g}\tilde{\rho}_{{\rm W}g})=0\ ;
\end{equation}
$|{\rm W}^g\rangle$ is manifestly distinguishable from
$|\tilde{{\rm W}}^g\rangle$, so the indistinguishability
$I(\rho_{{\rm W}g},\tilde{\rho}_{{\rm W}g})$ is clearly zero as
well, in accordance with Eq. (14) with $M(\rho_{Wg})=0$. Since
this three-qubit state is pure, this is an instance of Eq. (19):
the uniquely three-qubit entanglement, $\tau_{(3)}$, is seen to be
zero. Furthermore, for the two-qubit subsystems of this
three-qubit system, which are in mixed states, the entanglement
measure for $n=2$, $S_{(2)}^2$, calculated using their two-qubit
reduced states, $\rho_{(AB)}, \rho_{(BC)}, \rho_{(AC)}$, is
\begin{eqnarray}
S_{(AB)}^2(\rho_{(AB)})&=&4|\alpha|^2|\beta|^2=C^2(\rho_{AB})\nonumber\\
S_{(BC)}^2(\rho_{(BC)})&=&4|\beta|^2|\gamma|^2=C^2(\rho_{BC})\nonumber\\
S_{(AC)}^2(\rho_{(AC)})&=&4|\alpha|^2|\gamma|^2=C^2(\rho_{AB})\ ,
\end{eqnarray}
each taking the value ${4\over 9}$ for the traditional $W$ state
(that is, the state of Eq. (28) with
$\alpha=\beta=\gamma=\sqrt{\frac{1}{3}}$).

On the other hand, for generalized GHZ (or 3-cat) states,
\begin{equation}
|{\rm
GHZ}^g\rangle=\alpha\,|000\rangle+\sqrt{1-\alpha^2}\,|111\rangle\
,
\end{equation}
we find the entanglement behaves oppositely, that is
\begin{equation}
S_{(3)}^2={\rm Tr}(\rho_{{\rm GHZ}g}\tilde{\rho}_{{\rm
GHZ}g})=4|\alpha|^2(1-|\alpha|^2) =\tau_{(3)}\ ,
\end{equation}
(from Eq. 16, and analogously  for 2-cat states) while
$C^2_{(AB)}=C^2_{(BC)}=C^2_{(AC)}=0$ for all values of $\alpha$
and $S_{(3)}^2=\tau_{(3)}$ can reach the maximum value, $1$ (for
the three-cat state with $\alpha=\textstyle{\sqrt{\frac{1}{2}}}$,
see Eq. 25).

Now consider the following class of {\it mixed} three-qubit states
\begin{equation}
\rho_{m3}=w P[|000\rangle]+(1-w)P[|111\rangle]\ .
\end{equation}
with $w\in [0,1].$ Being mixed, these states obey the general
relation, Eq. (14) rather than its special case, Eq. (15). For
these states the mixedness is
\begin{equation}
M(\rho_{m3})=1-{\rm Tr}(\rho_{m3}^2)=2w(1-w)\ ,
\end{equation}
and the multipartite entanglement measure takes the value
\begin{equation}
S_{(3)}^2(\rho_{m3})={\rm Tr}(\rho_{m3}\tilde{\rho}_{m3})=2w(1-w)\
,
\end{equation}
as well. In accordance with Eq. (14), we also have
\begin{equation}
I(\rho_{m3},\tilde{\rho}_{m3})=4w(1-w)\ .
\end{equation}
In this case, we see that both $M$ and $S_{(3)}^2$ vary
proportionally to the degree of spin-flip symmetry.

Note that the states $\rho_{m3}$ can be viewed as three-qubit
reduced states arrived at by partial tracing out one of the qubits
from the class of generalized four-qubit Schr\"odinger cat pure
states (taking $\alpha^2\equiv w$),
\begin{equation}
|4{\rm cat}^g\rangle =\alpha\,|0000\rangle
+\sqrt{1-\alpha^2}\,|1111\rangle\ ,
\end{equation} with $\alpha\in
[0,1]$, for which the entanglement measure $S_{(4)}^2$ takes the
values
\begin{equation}
S_{(4)}^2={\rm Tr}(\rho_{4{\rm cat}g}\tilde{\rho}_{4{\rm
cat}g})=4\alpha^2(1-\alpha^2)\ .
\end{equation}
Because these states are pure, the mixedness $M(\rho)$ is zero; we
have that the four-qubit entanglement is the degree of spin-flip
symmetry $S_{(4)}^2=\tau_{(4)}(\rho_{4{\rm cat}g})=I(\rho_{4{\rm
cat}g},\tilde{\rho}_{4{\rm cat}g})$, and that
$S_{(3)}^2(\rho_{m3})=S_{(4)}^2(\rho_{4{\rm
cat}g})/2=\frac{1}{2}\tau_4$ is half that value, as is the
three-qubit mixedness.

\section{Conclusions}

By considering multipartite entanglement and mixedness together
with the degree of symmetry of quantum states under the $n$-qubit
spin-flip transformation, a general relation was found between
these fundamental properties.  Multipartite entanglement, as
described by a recently introduced measure, and state mixedness
were seen to be complementary within classes of states possessing
the same degree of spin-flip symmetry. For pure states, the value
of this multipartite entanglement measure, the degree of spin-flip
symmetry, and the $n$-tangle were seen to coincide. These results
can be expected to be particularly useful for the study of quantum
states used for quantum information processing in the presence of
decoherence.

\section{Acknowledgment}

We are grateful to the National Science Foundation (NSF); the
Center for Subsurface Imaging Systems (CenSSIS), an NSF
Engineering Research Center; and the Defense Advanced Research
Projects Agency (DARPA).

\vfill\eject

\end{document}